\documentclass[%
twocolumn,
superscriptaddress,
nofootinbib,
longbibliography,
aps,
pra
]{revtex4-2}

\usepackage{amsmath,amssymb,bm,braket,url,dsfont,xspace,enumitem}
\usepackage{graphicx}
\usepackage{xcolor}
\usepackage{array}
\usepackage{booktabs}

\usepackage{longtable}

\newcolumntype{L}{>{$}l<{$}} 
\newcolumntype{C}{>{$}c<{$}} 

\usepackage[
draft=false,
bookmarks=true,
bookmarksnumbered=false,
bookmarksopen=false,
bookmarksopenlevel=4,
colorlinks=true,
anchorcolor=black,
citecolor=blue,
filecolor=magenta,
linkcolor=blue,
linkbordercolor={1 0 0},
urlcolor=blue,
pdfborder={0 0 1},
pdftitle={},
pdfauthor={},
pdfsubject={},
pdfkeywords={},
pdfpagemode=UseThumbs,
]{hyperref}

\usepackage{orcidlink}

\newcommand{\bk}{{\bm{k}}}

\newcommand{\T}{$\theta$\xspace}
\newcommand{\PT}{$\theta I$\xspace}
\newcommand{\Pa}{$I$\xspace}

\newcommand{\sca}{\mathcal{S}}
\newcommand{\scafluc}{\mathcal{O}}
\newcommand{\system}{\mathcal{O}}

\begin{document}
\title{
Symmetry analysis of cross-circular and parallel-circular Raman optical activity
}

\author{Hikaru Watanabe
        \orcidlink{0000-0001-7329-9638}
        } 
\email{hikaru-watanabe@g.ecc.u-tokyo.ac.jp}
\affiliation{Department of Physics, University of Tokyo, Tokyo 113-0033, Japan}

\author{Rikuto Oiwa
        \orcidlink{0000-0001-6429-6136}
        } 
\affiliation{Department of Physics, Hokkaido University, {Sapporo}, Hokkaido 113-8656, Japan}
\affiliation{Center for Emergent Matter Science, RIKEN, {Wako}, Saitama 351-0198, Japan}~~

\author{Gakuto Kusuno
        } 
\affiliation{Department of Physics, Institute of Science Tokyo, Tokyo 152-8551, Japan}

\author{Takuya Satoh
        \orcidlink{0000-0001-6270-0617}
        } 
\affiliation{Department of Physics, Institute of Science Tokyo, Tokyo 152-8551, Japan}
\affiliation{Quantum Research Center for Chirality, Institute for Molecular Science, Aichi 444-8585, Japan}

\author{Ryotaro Arita
        \orcidlink{0000-0001-5725-072X}
        } 
\affiliation{Department of Physics, University of Tokyo, Tokyo 113-0033, Japan}
\affiliation{Center for Emergent Matter Science, RIKEN, {Wako}, Saitama 351-0198, Japan}

\begin{abstract}
        The Raman scattering regarding the circularly-polarized incident and scattered lights is closely related to the circular activity of a given system.
        We investigate the symmetry of its activity, called the cross-circular and parallel-circular Raman optical activity.
        The analysis is systematically performed with the magnetic point groups and indicates that the response allows for a useful diagnosis of the symmetry of materials like chirality and (magneto-)axiality.
        It is also shown that the Stokes and anti-Stokes processes are related to each other by the conserved antiunitary symmetry for the time-reversal operation and that combined with the mirror reflection.  
\end{abstract}

\maketitle

\section{introduction}
\label{Sec_introduction}

Raman scattering, light scattering with a slight frequency shift, is one of the powerful probes of materials.
The response serves to characterize elementary excitations like magnon and exciton and has been applied to a broad range of fields~\cite{Loudon2001-zi,Yu2010-as}.
For instance, its advantageous feature is evident that the Raman spectroscopy of magnetic materials points to the spectral weight of state and electrical activity of magnetic excitations~\cite{Greene1965-td,Tanabe1965-dx,Fleury1968-le}.
Recent studies further revealed that spectroscopy gives a deeper insight into spontaneous symmetry breaking when combined with the control of polarization of light.
It is shown that the Raman spectroscopy is convenient to investigate the order parameter and its coupling to the structural and electronic properties of micro-scaled samples such as van der Waals materials by the use of circularly-polarized and crossed linearly-polarized lights~\cite{Jin2018-zj,Jin2020-jz,McCreary2020-sj,Zhang2020-dg,Huang2020-hq,Lujan2022-of}.
In particular, the cross-circular and parallel-circular Raman spectroscopy of our focus is an established method for corroborating the phase of matter.

\expandafter\ifx\csname iffigure\endcsname\relax
                \begin{figure}[htbp]
                \centering
                \includegraphics[width=0.8\linewidth,clip]{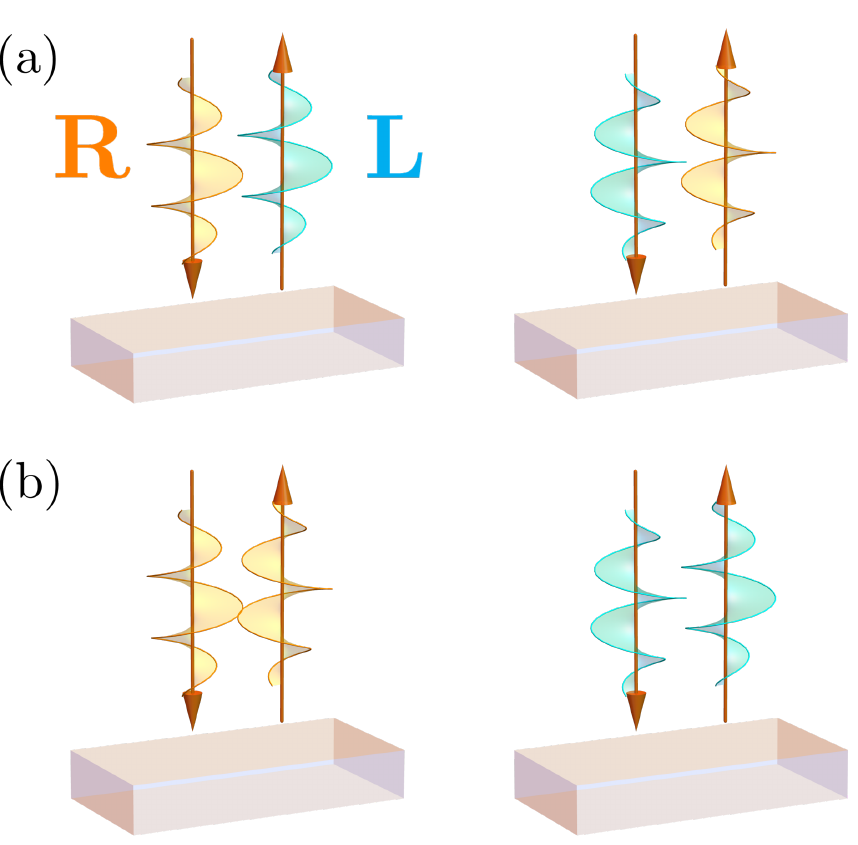}
                \caption{
                        Experimental setup for measuring (a) the cross-circular Raman optical activity and (b) the parallel-circular Raman optical activity.
                        The orange-colored arrows denote incident and reflected lights.
                        The circular polarization is defined by the rotation direction of the photo-electric field, colored in orange and cyan colors for right-handed and left-handed polarizations, in a plane fixed in the observatory frame.
                        The optical activities are defined by the difference in the measured quantities between the left and right panels in each of (a) and (b).  
                }
                \label{Fig_ROA-config}
                \end{figure}
\fi

Figure~\ref{Fig_ROA-config} depicts the setup of the cross-circular and parallel-circular Raman spectroscopy in the backward scattering geometry.
The activity, namely cross- and parallel-circular Raman optical activities (CCROA, PCROA), is found in ferromagnetic materials~\cite{Koshizuka1980-zz,Huang2020-hq, Cenker2021-bd}, nonmagnetic but mirror-asymmetric systems~\cite{Lacinska2022-ti,Yang2022-dl,Liu2023-qm,Zhao2023-ay,Martinez2024-nd}, and those with chiral structures~\cite{Ishito2022-fr,Ishito2023-lv,Oishi2024-dy}~\footnote{
        The parallel-circular and cross-circular optical activities are also called the activities regarding in-phase dual circular polarization and out-of-phase dual circular polarization, respectively~\cite{Vargek1997-rv}.
}.
In light of the microscopic mechanism, these Raman optical activities (ROA), termed by dual-circular ROA~\cite{Nafie1989-hj}, are related to antisymmetric Raman scattering~\cite{Koningstein1968-ci,Mortensen1968-gn} and vibronic Raman optical activity~\cite{Barron1971-tk,Chiu1970-ee,Barron1975-il} stemming from the resonant particle-hole excitations and symmetry breaking. 

Only recently, the Raman spectroscopic study~\cite{Kusuno_experiment} clarified that the conserved time-reversal symmetry is tied to the relationship between the Stokes and anti-Stokes processes, which respectively refer to the emission and absorption of the elementary excitation during the light scattering.
The experimental evidence implies that the dual-circular Raman spectroscopy allows us to identify the symmetry, including that concerning the time-reversal operation.
However, the systematic symmetry analysis has not been presented.

In this work, we present the symmetry analysis of dual-circular ROA.
The classification based on magnetic point groups allows us to identify the materials with these ROA and further points out the importance of the antiunitary symmetry.
Although the prior studies tabulated the symmetry-adapted Raman response~\cite{Ovander1960-fo,Cracknell1969-or}, our analysis shows that the magnetic symmetry is useful to identify the relation between the Stokes and anti-Stokes processes (creation and annihilation of elementary excitations during the light scattering).
The classification is based on the magnetic Laue, achiral, and chiral classes, providing a diagnosis of quantum materials manifesting exotic symmetry breaking.

The outline is as follows.
Section~\ref{SecSub_raman_basics} is for introducing the setup of our interest and explaining the symmetry constraints on the Raman-scattering intensity.
In Sec.~\ref{Sec_even_parity_CCROA} and Sec.~\ref{Sec_odd_parity_CCROA}, we classify CCROA characteristics of centrosymmetric and noncentrosymmetric materials, respectively.
The classification can work in the case of PCROA as well.
We discuss the CCROA and PCROA of cubic systems in Sec.~\ref{Sec_cubic_ROA}, while the preceding sections are dedicated to the analysis of non-cubic crystals.
We summarize our work and comment on the implications drawn from this work in Sec .~\ref{Sec_summary}.

\section{Setup}
\label{SecSub_raman_basics}

We present the symmetry analysis of the (spontaneous) Raman scattering process.
Let the light be parametrized by $(\bm{e}_X, \omega_X, \bm{q}_X)$, where $\bm{e}$ is the polarization vector, $\omega$ is the frequency, and $\bm{q}$ is the wave vector ($X=\text{i}$ for the incident light and $X=\text{f}$ for the scattered light).
In the following, we consider the Raman scattering process concerning an elementary excitation $\phi (\delta \omega, \bk )$ such as phonon and magnon.
The scattering intensities are expressed by~\cite{Loudon2001-zi}
                \begin{equation}
                U \propto \left| \bm{e}_\text{f}^\dagger \hat{\chi}  \bm{e}_\text{i} \right|^2 n_\phi^\text{S} (\delta \omega, \bk; \scafluc ),
                \label{raman_intensity_stokes}
                \end{equation}
for the Stokes process ($\omega_\text{i} > \omega_\text{s}$) and 
                \begin{equation}
                U \propto \left| \bm{e}_\text{f}^\dagger \hat{\xi}   \bm{e}_\text{i} \right|^2 n_\phi^\text{AS} (\delta \omega, \bk; \scafluc ),
                \label{raman_intensity_antistokes}
                \end{equation}
for the anti-Stokes process ($\omega_\text{i} < \omega_\text{s}$).
Note that the energy and momentum conservations hold as $\omega_\text{i} = \omega_\text{f} \pm  \delta \omega$ and $\bm{q}_\text{i} = \bm{q}_\text{f} \pm  \bk$ for the Stokes and anti-Stokes processes. 
$\hat{\chi}$ and $\hat{\xi}$ are the nonlinear susceptibilities that describe the emission and absorption of $\phi$ during the light scattering process. 
$n_\phi^\text{S}$ and $n_\phi^\text{AS}$ are the power spectra denoting the thermal population of $\phi$.
The power spectra depend on $\scafluc$, that is, the property of the target material such as crystal structure and spontaneous order.
For simplicity, we neglected other prefactors such as refractive indices by considering small detuning $|\delta \omega|\ll |\omega|$. 
The symmetry of light-matter interaction is built into the nonlinear susceptibilities.
Let us consider $\hat{\chi}$ written with its parameter dependence as
    \begin{equation}
        \chi_{ab} (\omega_\text{i}, \bm{q}_\text{i}; \delta \omega, \bk; \sca) = \Braket{e_a^\text{f}, \omega_\text{f}, \bm{q}_\text{f}; \phi (\delta \omega, \bk) | \sca | e_b^\text{i}, \omega_\text{i}, \bm{q}_\text{i}}.
    \end{equation}
The operator $\sca = \sca (\scafluc)$ denotes the scattering process with its implicit dependence on $\scafluc$.

The polarization state of light is defined with the circular coordinates; i.e., $\bm{e}_\text{R} = (1, i,0)/\sqrt{2}$ and $\bm{e}_\text{L} = (1, -i,0)/\sqrt{2}$ for the right-handed and left-handed circularly-polarized lights.
As a result, the components of $\hat{\chi}$ are written down by such as $\chi_\text{RL}$.
The circular polarization is given according to the observatory frame, not by the helicity of light (see Fig.~\ref{Fig_ROA-config}).
Then, we define the cross-circular and parallel-circular Raman optical activities by the difference of the intensities between $(\bm{e}_\text{i},\bm{e}_\text{f}) = (\bm{e}_\text{R},\bm{e}_\text{L})$ and $(\bm{e}_\text{i},\bm{e}_\text{f}) = (\bm{e}_\text{L},\bm{e}_\text{R})$ in the cross-circular arrangement [Fig.~\ref{Fig_ROA-config}(a)], and that between $(\bm{e}_\text{i},\bm{e}_\text{f}) = (\bm{e}_\text{R},\bm{e}_\text{R})$ and $(\bm{e}_\text{i},\bm{e}_\text{f}) = (\bm{e}_\text{L},\bm{e}_\text{L})$ in the parallel-circular arrangement [Fig.~\ref{Fig_ROA-config}(b)], respectively.

Specifically, following reported experimental works, we consider the backward reflection geometry, where the light perpendicularly incident on the sample surface is antiparallel to the scattered light ($\bm{q}_\text{i} = q_i \hat{z}$, $\bm{q}_\text{f} = - q_i \hat{z}$).
We aim to identify the relation between dual-circular Raman optical activity and the bulk symmetry of a given material.
To eliminate spurious effects such as the birefringence effect and misalignment of the incident direction from the optical axes~\cite{Porto1966-wn,Hoffman1994-ik,Zhang2017-sh}, we assume that the light is propagating along the principal axis $\hat{z}$ of the system.
We also assume that the system is non-cubic, while we present the symmetry considerations for the cubic point groups later (Sec.~\ref{Sec_cubic_ROA})~\footnote{
        In the case of orthorhombic crystals, there is an ambiguity in choosing the axis from $\hat{x},\hat{y},\hat{z}$.
        We take the $\hat{z}$ axis in this work, but this specific choice does not affect the conclusion. 
}.

Let us consider the symmetry constraint on the nonlinear susceptibilities $\chi_{ab}$ and $\xi_{ab}$.
Since we are interested in the back-scattering geometry, it suffices to consider an operation $g$ that does not rotate the incident plane; \textit{i.e,} the symmetry constraints originate from the space-inversion ($g= I$), the $n$-fold rotation normal to the incident plane ($g= n$), the $2$-fold rotation parallel to the incident plane ($g= 2_\perp$), time-reversal operation ($g= \theta$), and combinations of afore-mentioned operations such as  $g= \theta I $.

First, we consider the space-inversion operation.
The susceptibility $\chi_{ab}$ for the Stokes process is transformed as
\begin{align}
&\chi_{ab} (\omega_\text{i}, \bm{q}_\text{i}; \delta \omega, \bk; \sca) \notag\\
    &=\Braket{e_a^\text{f}, \omega_\text{f}, \bm{q}_\text{f}; \phi (\delta \omega, \bk) | \sca | e_b^\text{i}, \omega_\text{i}, \bm{q}_\text{i}},\\
    &=\Braket{\hat{I}e_a^\text{f}, \omega_\text{f}, - \bm{q}_\text{f}; \hat{I} \phi (\delta \omega,\bk) | \hat{I} \sca \hat{I}^{-1} | \hat{I} e_b^\text{i}, \omega_\text{i}, - \bm{q}_\text{i}},\\
    &= \rho_I \Braket{e_a^\text{f}, \omega_\text{f}, - \bm{q}_\text{f}; \phi (\delta \omega,-\bk) | \hat{I} \sca \hat{I}^{-1} | e_b^\text{i}, \omega_\text{i}, - \bm{q}_\text{i}},\\
    &= \rho_I \chi_{ab} (\omega_\text{i}, -\bm{q}_\text{i}; \delta \omega, -\bk; \hat{I}\sca\hat{I}^{-1}),
    \label{nonlinear_susceptibility_I_constraint}
\end{align}
$\rho_I$ denotes the \Pa{} parity of $\phi$.
If $\hat{I} \sca \hat{I}^{-1} = \sca (\hat{I} \scafluc \hat{I}^{-1}) = \sca (\scafluc)$ holds due to the \Pa{} symmetry of the system, the scattering event is nonreciprocal, as is evident from the relation  
    \begin{equation}
        \left| \bm{e}_\text{f}^\dagger \hat{\chi} (\omega_\text{i}, \bm{q}_\text{i}; \delta \omega, \bk; \sca)  \bm{e}_\text{i} \right|^2 = \left| \bm{e}_\text{f}^\dagger \hat{\chi} (\omega_\text{i}, -\bm{q}_\text{i}; \delta \omega, -\bk; \sca)  \bm{e}_\text{i} \right|^2,
    \end{equation}
where the right-hand side denotes the scattering event whose experimental arrangement is the space-inversion (\Pa{}) image of the original configuration.
The total scattering intensity [Eq.~\eqref{raman_intensity_stokes}] is also nonreciprocal because the \Pa{} symmetry of $\scafluc$ leads to
                \begin{equation}
                        n_\phi^\text{S} (\delta \omega, \bk ; \scafluc)  = n_\phi^\text{S} (\delta \omega, -\bk; \hat{I}\scafluc \hat{I}^{-1})= n_\phi^\text{S} (\delta \omega, -\bk; \scafluc).
                \end{equation}

The purely nonreciprocal property does not hold in \Pa{}-broken systems.
One can decompose the scattering intensity into the nonreciprocal and reciprocal parts as
        \begin{widetext}
                \begin{align}
                        \left| \bm{e}_\text{f}^\dagger \hat{\chi} (\omega_\text{i}, \bm{q}_\text{i}; \delta \omega, \bk; \sca)  \bm{e}_\text{i} \right|^2 n_\phi^\text{S} (\delta \omega, \bk; \scafluc)
                                 &= \frac{1}{2} \left\{ \left| \bm{e}_\text{f}^\dagger \hat{\chi} (\omega_\text{i}, \bm{q}_\text{i}; \delta \omega, \bk; \sca)  \bm{e}_\text{i} \right|^2 n_\phi^\text{S} (\delta \omega, \bk; \scafluc) + \left| \bm{e}_\text{f}^\dagger \hat{\chi} (\omega_\text{i}, -\bm{q}_\text{i}; \delta \omega, -\bk; \sca)  \bm{e}_\text{i} \right|^2 n_\phi^\text{S} (\delta \omega, -\bk; \scafluc) \right\} \notag \\
                                 &+ \frac{1}{2} \left\{ \left| \bm{e}_\text{f}^\dagger \hat{\chi} (\omega_\text{i}, \bm{q}_\text{i}; \delta \omega, \bk; \sca)  \bm{e}_\text{i} \right|^2 n_\phi^\text{S} (\delta \omega, \bk; \scafluc) - \left| \bm{e}_\text{f}^\dagger \hat{\chi} (\omega_\text{i}, -\bm{q}_\text{i}; \delta \omega, -\bk; \sca)  \bm{e}_\text{i} \right|^2 n_\phi^\text{S} (\delta \omega, -\bk; \scafluc) \right\}.
                \end{align}
        \end{widetext}
The first line is nonreciprocal, and the second line is the reciprocal contribution arising from the \Pa{} symmetry breaking.
By using the \Pa{}-parity characterization, we define the reciprocal and nonreciprocal dual-circular ROA.
For instance, when CCROA of the Stokes process ($\omega_\text{i} >\omega_\text{s}$) is given by 
        \begin{widetext}
                \begin{equation}
                \text{CC} (\omega_\text{i}, \bm{q}_\text{i}; \delta \omega, \bk; \system)    \equiv   \left\{  \left| \chi_\text{LR} (\omega_\text{i}, \bm{q}_\text{i}; \delta \omega, \bk; \sca)  \right|^2  -   \left| \chi_\text{RL} (\omega_\text{i}, \bm{q}_\text{i}; \delta \omega, \bk; \sca)  \right|^2 \right\} n_\phi^\text{S} (\delta \omega, \bk; \scafluc),
                \end{equation}
        \end{widetext}
the nonreciprocal and even-parity contribution is given by 
                \begin{align}
                &\text{CC}_+ (\omega_\text{i}, \delta \omega, \system) \notag \\
                &\equiv \frac{1}{2} \left\{ \text{CC} (\omega_\text{i}, \bm{q}_\text{i}; \delta \omega, \bk; \system)  + \text{CC} (\omega_\text{i}, -\bm{q}_\text{i}; \delta \omega, -\bk; \system)  \right\},
                \label{CCROA_nonreciprocal_def}
                \end{align}
while reciprocal and odd-parity part is 
        \begin{align}
        &\text{CC}_- (\omega_\text{i}, \delta \omega, \system) \notag \\
        &\equiv \frac{1}{2} \left\{ \text{CC} (\omega_\text{i}, \bm{q}_\text{i}; \delta \omega, \bk; \system)  - \text{CC} (\omega_\text{i}, -\bm{q}_\text{i}; \delta \omega, -\bk; \system)  \right\}.
        \label{CCROA_reciprocal_def}
        \end{align}

PCROA of the Stokes process is defined by
\begin{widetext}
        \begin{equation}
        \text{PC} (\omega_\text{i}, \bm{q}_\text{i}; \delta \omega, \bk; \system)    \equiv    \left\{ \left| \chi_\text{RR} (\omega_\text{i}, q_i \hat{z}; \delta \omega, \bk; \sca)   \right|^2  -   \left| \chi_\text{LL} (\omega_\text{i}, q_i \hat{z}; \delta \omega, \bk; \sca)   \right|^2  \right\} n_\phi^\text{S} (\delta \omega, \bk; \scafluc),
        \end{equation}
\end{widetext}
and its even-parity (PC$_+$) and odd-parity (PC$_-$) terms are obtained in a similar manner to the case of CCROA.
Note that the odd-parity dual-circular ROA (CC$_-$ and PC$_-$) vanishes if $\hat{I}\system \hat{I}^{-1} = \system$.

The symmetry analysis of $\hat{\chi}$ is similarly performed in the case of another unitary operation.
The important consequence is drawn from the mirror operation $m_\perp$, that is, the mirror reflection with respect to the incident plane of light.
The operation interchanges the circular polarization of light as $\hat{m}_\perp \bm{e}_\text{R} = \bm{e}_\text{L}$,~ $\hat{m}_\perp \bm{e}_\text{L} = \bm{e}_\text{R}$.
Thus, if the system is $m_\perp$-symmetric, dual-circular ROA is forbidden; e.g., CCROA vanishes as  
                \begin{equation}
                        \text{CC}_X (\omega_\text{i}, \delta \omega, \system) = - \text{CC}_X (\omega_\text{i}, \delta \omega, \system) = 0,
                        \label{CCROA-out_of_plane_mirror_constraint}
                \end{equation}
for $X= \pm$.
As a result, the systems of our interest are what lack the $m_\perp$ symmetry.
The $m_\perp$ breaking occurs with or without the simultaneous \Pa{} symmetry breaking which will be discussed in Sec.~\ref{Sec_even_parity_CCROA} and Sec.~\ref{Sec_odd_parity_CCROA}, respectively.

Next, we consider the time-reversal (\T{}) operation.
The nonlinear susceptibility $\chi_{ab}$ undergoes the transformation as
        \begin{align}
        &\chi_{ab} (\omega_\text{i}, \bm{q}_\text{i}; \delta \omega, \bk; \sca) \notag\\
            &=\Braket{\hat{\theta} e_b^\text{i}, \omega_\text{i}, - \bm{q}_\text{i}  | \hat{\theta} \sca^\dagger \hat{\theta}^{-1} | \hat{\theta}e_a^\text{f}, \omega_\text{f}, - \bm{q}_\text{f}; \hat{\theta} \phi (\delta \omega,\bk)},\\
            &=  \Braket{ e_{\overline{b}}^\text{i}, \omega_\text{i}, - \bm{q}_\text{i}  | \overline{\sca} | e_{\overline{a}}^\text{f}, \omega_\text{f}, - \bm{q}_\text{f};\overline{\phi} (\delta \omega,-\bk)},\\
            &= \xi_{\overline{b}\overline{a}} (\omega_\text{f}, \bm{q}_\text{f}; \delta \omega, \bk; \overline{\sca}),
        \end{align}
where $e_{\overline{b}} \equiv \theta e_{b}$, $\overline{\sca} \equiv \hat{\theta} \sca^\dagger \hat{\theta}^{-1}$, and $\overline{\phi} = \theta \phi$.
The operator $\overline{\sca}$, that is the \T{}-partner of $\sca$, does not violate the causality but shows the opposite polarity in its time-reversal-odd properties such as magnetization~\cite{Naguleswaran1998-yu}.
Following Ref.~\cite{Loudon1978-ud}, it is shown that the power spectra of $\phi$ and $\overline{\phi}$ are related as 
                \begin{equation}
                        f_\text{B} (\delta \omega ) n_\phi^\text{S} (\delta \omega, \bk ; \scafluc) = \left\{ 1 + f_\text{B} (\delta \omega )\right\} n_{\overline{\phi}}^\text{AS} (\delta \omega, -\bk ; \overline{\scafluc}),
                \end{equation}
with the Bose-Einstein distribution function $f_\text{B} (\delta \omega )$ and $\overline{\scafluc} = \hat{\theta} \scafluc \hat{\theta}^{-1}$.
As a result, the \T{} operation relates the Stokes and anti-Stokes Raman scattering intensities whose frequency of incident light are $\omega_\text{i}$ and $\omega_\text{f} = \omega_\text{i} - \delta \omega $, as clarified in Refs.~\cite{Loudon1978-ud,Yu2010-as}.
In terms of CCROA, considering $e_\text{R} = \hat{\theta} e_\text{L}$ and $e_\text{L} = \hat{\theta} e_\text{R}$, we obtain the relation 
                \begin{equation}
                \text{CC} (\omega_\text{i}, \bm{q}_\text{i}; \delta \omega, \bk; \system) = \frac{1 + f_\text{B} (\delta \omega )}{f_\text{B} (\delta \omega )} \text{CC} (\omega_\text{f}, -\bm{q}_\text{f}; -\delta \omega, -\bk; \overline{\system}),
                \label{CCROA_time-reversal}
                \end{equation}
where CCROA with $-\delta \omega <0$ (right-hand side) is defined by the anti-Stokes scattering intensities as
\begin{widetext}
        \begin{equation}
        \text{CC} (\omega_\text{f}, -\bm{q}_\text{f}; -\delta \omega, -\bk; \overline{\system})  = \left\{ \left| \xi_\text{LR} (\omega_\text{f}, -\bm{q}_\text{f}; \delta \omega, -\bk; \overline{\sca}) \right|^2 -   \left| \xi_\text{RL} (\omega_\text{f}, -\bm{q}_\text{f}; \delta \omega, -\bk; \overline{\sca}) \right|^2 \right\} n_{\overline{\phi}}^\text{AS} (\delta \omega, -\bk; \overline{\scafluc}).
        \end{equation}
\end{widetext}
Since the detuning frequency $\delta \omega = |\omega_\text{i} - \omega_\text{f}|$ is much smaller than the frequency of light $\omega_\text{i}$, we may approximate Eq.~\eqref{CCROA_time-reversal} as
                \begin{equation}
                \text{CC} (\omega_\text{i}, \bm{q}_\text{i}; \delta \omega, \bk; \system) \approx \frac{1 + f_\text{B} (\delta \omega )}{f_\text{B} (\delta \omega )} \text{CC} (\omega_\text{i}, \bm{q}_\text{i}; -\delta \omega, -\bk; \overline{\system}),
                \label{CCROA_timereversal_approximate_relation}
                \end{equation}
which indicates the symmetry of CCROA in the Stokes and anti-Stokes processes whose systems are parametrized by $\system$ and $\overline{\system}$, respectively.
We also used $\bm{q}_\text{f} = - \bm{q}_\text{i}$.

If the system is \T{}-symmetric, $\overline{\system} = \system$ holds, resulting in the same signs between the stokes and anti-Stokes CCROA as
                \begin{equation}
                        \text{CC} (\omega_\text{i}, \bm{q}_\text{i}; \delta \omega, \bk; \system) \approx \frac{1 + f_\text{B} (\delta \omega )}{f_\text{B} (\delta \omega )} \text{CC} (\omega_\text{i}, \bm{q}_\text{i}; -\delta \omega, -\bk; \system).
                \end{equation}

Furthermore, the combined operation of $\theta$ and $m_\perp$ leads to the relation complementary to Eq.~\eqref{CCROA_time-reversal}.
In the case of CCROA, the following relation is obtained
        \begin{align}
        &\text{CC} (\omega_\text{i}, \bm{q}_\text{i}; \delta \omega, \bk; \system) \notag \\
        &= - \frac{1 + f_\text{B} (\delta \omega )}{f_\text{B} (\delta \omega )} \text{CC} (\omega_\text{f}, -\bm{q}_\text{f}; -\delta \omega, -\bk; \overline{\system}'),\\
        &\approx - \frac{1 + f_\text{B} (\delta \omega )}{f_\text{B} (\delta \omega )} \text{CC} (\omega_\text{i}, \bm{q}_\text{i}; -\delta \omega, -\bk; \overline{\system}'),
        \label{CCROA_time-reversal_mirror_combined}
        \end{align}
where $\overline{\system}' = \hat{\theta} \hat{m}_\perp \system^\dagger (\hat{\theta} \hat{m}_\perp)^{-1}$.
As a result, a $\theta m_\perp$-symmetric systems may show the Stokes and anti-Stokes CCROA with the opposite signs, which is in stark contrast to the \T{}-symmetric systems.
Note that the $\theta m_\perp$-symmetric part of CCROA is forbidden if the system also respect the \T{} symmetry because of the conserved $m_\perp$ symmetry (see Eq.~\eqref{CCROA-out_of_plane_mirror_constraint}).
Thus, the obtained contribution is unique to \T{}-broken systems.

To summarize, CCROA consists of four contributions as follows.
First, it is divided into the even- and odd-parity parts as
                \begin{equation}
                \text{CC}_+ (\omega_\text{i},\delta \omega ,\system) + \text{CC}_- (\omega_\text{i},\delta \omega ,\system).
                \end{equation}
Each part is further classified into the \T{}-even and \T{}-odd contributions which manifest the same (symmetric) and opposite (anti-symmetric) signs between the Stokes and anti-Stokes signals, respectively.
In light of the \T{} parity, CC$_X$ is decomposed as
                \begin{equation}
                        \text{CC}_X (\omega_\text{i},\delta \omega ,\system) = \text{CC}_X^\text{\,s} (\omega_\text{i},\delta \omega ,\system) + \text{CC}_X^\text{\,a} (\omega_\text{i},\delta \omega ,\system),
                \end{equation}
where CC$_+^\text{\,s}$ is \T{}-even ($\theta m_\perp$-odd) and CC$_+^\text{\,a}$ is \T{}-odd ($\theta m_\perp$-even).

Following the parallel discussions, PCROA is classified by the parity with respect to the \Pa{} and \T{} operations.
PCROA, however, differs from CCROA in terms of the Stokes and anti-Stokes symmetries ensured by the \T{} or $\theta m_\perp$ symmetry; in \T{}-symmetric systems, 
        \begin{align}
                &\text{PC} (\omega_\text{i}, \bm{q}_\text{i}; \delta \omega, \bk; \system) \notag \\
                &\approx - \frac{1 + f_\text{B} (\delta \omega )}{f_\text{B} (\delta \omega )} \text{PC} (\omega_\text{i}, \bm{q}_\text{i}; -\delta \omega, -\bk; \system),
        \end{align}
exhibiting the opposite signs between the emission and absorption processes.
On the other hand, if the $\theta m_\perp$ symmetry is intact, the following relation holds: 
        \begin{equation}
                \text{PC} (\omega_\text{i}, \bm{q}_\text{i}; \delta \omega, \bk; \system) \approx  \frac{1 + f_\text{B} (\delta \omega )}{f_\text{B} (\delta \omega )} \text{PC} (\omega_\text{i}, \bm{q}_\text{i}; -\delta \omega, -\bk; \system).
        \end{equation}
Then, the \T{}-even term is labeled by the anti-symmetric PCROA (PC$_X^\text{\,a}$), while the \T{}-odd term is by the symmetric part (PC$_X^\text{\,s}$).

As a result, the symmetry between the Stokes and anti-Stokes signals is contrasting between the \T{}-symmetric system and \T{}-violating but $\theta m_\perp$-conserving system.
It implies that careful observation of Stokes and anti-Stokes signals enables us to identify the antiunitary symmetry in materials.
Note that, when both of the \T{} and $\theta m_\perp$ symmetries are broken, both of the symmetric and anti-symmetric components are allowed.
Similar arguments are found in Ref.~\cite{Hecht1993-ag} investigating CCROA and PCROA of the chiral molecules as well as in discussions concerning another type of ROA for the odd-parity and \T{}-even~\cite{Barron1976-hs,Barron1985-mg} and the even-parity and \T{}-odd cases~\cite{Barron1982-nx}.
On the other hand, our symmetry analysis further generalizes their discussions to cover more diverse cases; e.g., the even-parity and \T{}-even or odd-parity and \T{}-odd dual-circular ROA, which are not realized in molecular gas and solution.
In the following sections, however, we generalize their results from the viewpoint of symmetry and classify all the magnetic point groups in terms of the \Pa{} and \T{} parties of CCROA and PCROA.

\section{Even-parity cross-circular Raman optical activity}
\label{Sec_even_parity_CCROA}

We consider the even-parity CCROA (CC$_+$) to delve into the symmetry of the Raman scattering process.
The $I$-operation constraint [Eq.~\eqref{nonlinear_susceptibility_I_constraint}] leads to the relation
                \begin{equation}
                        \text{CC}_+ (\omega,\delta \omega ,\system) = \text{CC}_+ (\omega, \delta \omega ,I \system I^{-1}).
                        \label{CCROA_even_inversion_constraint}
                \end{equation}
Thus, the even-parity part is identical between the original system ($\system$) and its inversion motif ($I \system I^{-1}$), following that the point group of the target material $\bm{G}$ shows the CC$_+$ symmetry same as that of the point group enhanced by the space-inversion operation as $\bm{G} \cup I \bm{G}$. 

The classification of CC$_+$ is therefore given by the Laue class, a series of the point groups that are merged into the same group after the inversion-operation enhancement.
For instance, the trigonal point groups are classified into the Laue classes $\bar{3}$ and $\bar{3}m$.
They are explicitly given as
                \begin{equation*}
                \bar{3} = \left\{ 3, \bar{3}\right\},~\bar{3}m = \left\{ 32, 3m, \bar{3}m \right\}.
                \label{laue_class_trigonal_monocolor}
                \end{equation*}
The former class does not show the $m_\perp$ symmetry, while the latter does.
As a result, the even-parity CCROA is allowed in the $\bar{3}$-class materials, while it is forbidden in the $\bar{3}m$-class materials.

Furthermore, Eqs.~\eqref{CCROA_timereversal_approximate_relation} and \eqref{CCROA_time-reversal_mirror_combined} imply that the antiunitary operation relates the CC$_+$ of the Stokes process with that of the anti-Stokes process.
Thus, it is convenient to classify the response in terms of the magnetic point groups.
122 magnetic point groups are classified into the magnetic-group analog of the Laue class, that is magnetic Laue class.
Similarly to the magnetic point groups, the magnetic Laue class is defined by the series of magnetic point groups merged by the space-inversion operation.
The magnetic Laue classes are classified into three types as in the case of magnetic point groups: colorless, gray, and black-white Laue classes.
The colorless class does not have any antiunitary element like $\theta$.

The gray and black-white Laue classes for trigonal systems are obtained as
        \begin{align*}
        &\bar{3}1' = \left\{ 31', \bar{3}1', \bar{3}'\right\},\\
        &\bar{3}m1' = \left\{ 321', 3m1', \bar{3}m1', \bar{3}'m, \bar{3}m' \right\},\\
        &\bar{3}m' = \left\{ 32', 3m', \bar{3}m'\right\}.
        \end{align*}
Note that the gray Laue class can cover black-white point groups.
For instance, the magnetic point group $\bar{3}'$ is included in the gray Laue class $\bar{3}1'$.
We obtained the five magnetic Laue classes for the trigonal system in total.
Owing to the $m_\perp$ symmetry, the classes $\bar{3}m$ and $\bar{3}m1'$ do not have the activity concerning CC$_+$, while the rest of the classes ($\bar{3}$, $\bar{3}1'$, $\bar{3}m'$) do.

For the gray Laue class $\bar{3}1'$, CCROA is attributed to CC$_+^\text{\,s}$ because of the $\theta$ symmetry.
Materials belonging to this class are found in a series of so-called ferroaxial (ferro-rotational) materials where the ferroaxial vector $\bm{A}$, a \T{}-even axial vector implying the $m_\perp$-symmetry violation, can be present due to their crystal structures or structural phase transition~\cite{Johnson2011-he,Hlinka2016-oo,Hayami2022-yq,Yamagishi2023-qm,Bhowal2024-pw,He2024-ie,Day-Roberts2025-ga}.  
Candidate materials undergoing the ferroaxial phase transition include RbFe(MoO$_4$)$_2$~\cite{Kenzelmann2007-bw,Jin2020-xk,Yamagishi2023-qm}, K$_2$Zr(PO$_4$)$_2$~\cite{Yamagishi2023-qm, Bhowal2024-pw}, NiTiO$_3$~\cite{Hayashida2020-tu}, MnTiO$_3$~\cite{Sekine2024-di,Zhang2024-ef}, Ca$_5$Ir$_3$O$_{12}$~\cite{Hanate2021-ai,Hanate2023-ht,Hayami2023-nu}, Na$_2$Ba$M$(PO$_4$)$_2$~\cite{Kajita2024-wu,Kajita2025-ip}, van der Waals materials in the charge-density-wave phase~\cite{Lacinska2022-ti,Song2022-xr,Yang2022-dl,Liu2023-qm,Zhao2023-ay}, and so on.
Being consistent with the symmetry analysis, the same sign of the CCROA signals between the Stokes and anti-Stokes signals has been observed~\cite{Kusuno_experiment}.  

On the other hand, the black-white Laue class $\bar{3}m'$ allows for the \T{}-odd and antisymmetric part of the even-parity CCROA denoted by CC$_+^\text{\,a}$.
The response is not admixed with the \T{}-even counterpart CC$_+^\text{\,s}$ because of the $\theta m_\perp$ symmetry.
The black-white Laue class shows the magneto-axial symmetry whose unitary symmetry is the same as that of the ferroaxial system, but \T{} symmetry is not kept without combining with the $m_\perp$ or $2_\perp$ operation.
The magneto-axial symmetry is found in ferromagnetic systems.
Experimental observations have been made in ferromagnets~\cite{Koshizuka1980-zz,Huang2020-hq,Cenker2021-bd} and systems under the external magnetic field~\cite{Sirenko1997-oy,Kossacki2012-ft}.
Note that the activity can be present in not only the ferromagnets but also the antiferromagnets manifesting the magneto-axial symmetry like Mn$_3$Sn~\cite{Nakatsuji2022-vr,Smejkal2022-rk}.
Both of the \T{}-even and \T{}-odd CC$_+$ activities coexist in the colorless Laue class $\bar{3}$.

The classification based on the (non-cubic) magnetic Laue class is completed in Table~\ref{Table_magnetic-Laue-class}.
Some of the black-white Laue classes allow for both of CC$_+^\text{\,s}$ and CC$_+^\text{\,a}$ because of the ferroaxial motif of their crystal structure.

\begin{longtable}{ccccc}
        \caption{
                Classification of the even-parity cross-circular Raman optical activity (CC$_+$) by magnetic Laue class.
                Each class is labeled by the \T{}-even and \T{}-odd CCROA; \textit{e.g.}, $(\times, \checkmark)$ denotes no \T{}-even CCROA but allowed \T{}-odd CCROA.
                The magnetic Laue class is comprised of magnetic point groups ($M=G$ or $M = G \cup \theta g G$) in which unitary operations $g$ form the group $G$ and antiunitary operations are given by $\theta g \, G$.
                CCROA with the superscript `$\ast$' denotes what may be admixed with the birefringence effect.
                }
        \label{Table_magnetic-Laue-class}\\
        
        \toprule
        Laue class & CCROA & $M$ & $G$ & $g$ \\
        \midrule
        \endfirsthead
        
        \multicolumn{5}{c}{\tablename\ \thetable\ (\textit{cont.})} \\
        Laue class & CCROA & $M$ & $G$ & $g$ \\
        \midrule
        \endhead
        
        \endfoot
        
        \endlastfoot
        \multicolumn{5}{l}{(colorless Laue class)}\\
                 $\bar{1}$ & $(\checkmark,\checkmark)^\ast$ & $1$ & $1$& \\ 
                  & & $\bar{1}$ & $\bar{1}$& \\ 
                 $2/m$ & $(\checkmark,\checkmark)^\ast$ & $2$ & $2$& \\ 
                 & & $m$ & $m$& \\ 
                 & & $2/m$ & $2/m$& \\ 
                 $mmm$ & $(\times,\times)$ & $2mm$ & $2mm$& \\ 
                 & & $222$ & $222$& \\ 
                 & & $mmm$ & $mmm$& \\ 
                 $\bar{3}$ & $(\checkmark,\checkmark)$ & $3$ & $3$& \\ 
                 & & $\bar{3}$ & $\bar{3}$& \\ 
                 $\bar{3}m$ & $(\times,\times)$ & $32$ & $32$& \\ 
                 & & $3m$ & $3m$& \\ 
                 & & $\bar{3}m$ & $\bar{3}m$& \\ 
                 $4/m$ & $(\checkmark,\checkmark)$ & $4$ & $4$& \\ 
                 & & $\bar{4}$ & $\bar{4}$& \\ 
                 & & $4/m$ & $4/m$& \\ 
                 $6/m$ & $(\checkmark,\checkmark)$ & $6$ & $6$& \\ 
                 & & $\bar{6}$ & $\bar{6}$& \\ 
                 & & $6/m$ & $6/m$& \\ 
                 $4/mmmm$ & $(\times,\times)$ & $422$ & $422$& \\ 
                 & & $4mm$ & $4mm$& \\ 
                 & & $\bar{4}2m$ & $\bar{4}2m$& \\ 
                 & & $4/mmm$ & $4/mmm$& \\ 
                 $6/mmmm$ & $(\times,\times)$ & $622$ & $622$& \\ 
                 & & $6mm$ & $6mm$& \\ 
                 & & $\bar{6}2m$ & $\bar{6}2m$& \\ 
                 & & $6/mmm$ & $6/mmm$& \\ 
                \midrule
        \multicolumn{5}{l}{(gray and black-white Laue classes)}\\
                $\bar{1}1'$ & $(\checkmark,\times)^\ast$ & $1'$ & $1$&$1$ \\ 
                  & & $\bar{1}1'$ & $\bar{1}$& $1$ \\ 
                  & & $\bar{1}'$ & $1$& $\bar{1}$ \\ 
                 $2/m1'$ & $(\checkmark,\times)^\ast$ & $21'$ & $2$& $1$\\ 
                 & & $m1'$ & $m$&$1$ \\ 
                 & & $2/m1'$ & $2/m$&$1$ \\ 
                 & & $2/m'$ & $2$&$m$ \\ 
                 & & $2'/m$ & $m$&$2$ \\ 
                 $2'/m'$ & $(\checkmark,\checkmark)^\ast$ & $m'$ & $1$& $m$\\ 
                 & & $2'$ & $1$&$2$ \\ 
                 & & $2'/m'$ & $\bar{1}$&$2$ \\ 
                 $mmm1'$ & $(\times,\times)$ & $2mm1'$ & $2mm$&$1$ \\ 
                 & & $2221'$ & $222$& $1$\\ 
                 & & $mmm1'$ & $mmm$& $1$\\ 
                 & & $m'm'm'$ & $222$& $\bar{1}$\\ 
                 & & $m'mm$ & $2mm$& $\bar{1}$\\ 
                 $m'mm$ & $(\times ,\checkmark)$ & $2'2'2$ & $2$&$1$ \\ 
                 & & $m'm'2$ & $2$& $m$\\ 
                 & & $m'm2'$ & $m$& $2$\\ 
                 & & $m'mm$ & $2/m$& $2$\\ 
                 $\bar{3}1'$ & $(\checkmark,\times)$ & $31'$ & $3$& $1$\\ 
                 & & $\bar{3}1'$ & $\bar{3}$& $1$\\ 
                 & & $\bar{3}'$ & $3$& $\bar{1}$\\ 
                 $\bar{3}m1'$ & $(\times,\times)$ & $321'$ & $32$& $1$\\ 
                 & & $3m1'$ & $3m$& $1$\\ 
                 & & $\bar{3}m1'$ & $\bar{3}m$& $1$\\ 
                 & & $\bar{3}'m$ & $3m$& $\bar{1}$\\ 
                 & & $\bar{3}'m'$ & $32$& $\bar{1}$\\ 
                 $\bar{3}m'$ & $(\times,\checkmark)$ & $32'$ & $3$& $2_\perp$\\ 
                 & & $3m'$ & $3$& $m_\perp$\\ 
                 & & $\bar{3}m'$ & $\bar{3}$& $m_\perp$\\ 
                 $4/m1'$ & $(\checkmark,\times)$ & $41'$ & $4$& $1$\\ 
                 & & $\bar{4}1'$ & $\bar{4}$& $1$\\ 
                 & & $4/m1'$ & $4/m$& $1$\\ 
                 & & $4/m'$ & $4$& $\bar{1}$\\ 
                 & & $4'/m'$ & $\bar{4}$& $\bar{1}$\\ 
                 $4'/m$ & $(\checkmark,\checkmark)$ & $4'$ & $2$& $4$\\ 
                 & & $\bar{4}'$ & $2$& $\bar{4}$\\ 
                 & & $4'/m$ & $2/m$& $\bar{4}$\\ 
                 $6/m1'$ & $(\checkmark,\times)$ & $61'$ & $6$& $1$\\ 
                 & & $\bar{6}1'$ & $\bar{6}$& $1$\\ 
                 & & $6/m1'$ & $6/m$& $1$\\ 
                 & & $6/m'$ & $6$& $\bar{1}$\\ 
                 & & $6'/m$ & $\bar{6}$& $\bar{1}$\\ 
                 $6'/m'$ & $(\checkmark,\checkmark)$ & $\bar{6}'$ & $3$& $m$\\ 
                 & & $6'$ & $3$& $2$\\ 
                 & & $6'/m'$ & $\bar{3}$& $2$\\ 
                 $4/mmm1'$ & $(\times,\times)$ & $4221'$ & $422$& $1$\\ 
                 & & $4mm1'$ & $4mm$& $1$\\ 
                 & & $\bar{4}2m1'$ & $\bar{4}2m$&$1$ \\ 
                 & & $4/mmm1'$ & $4/mmm$& $1$ \\ 
                 & & $4/m'm'm'$ & $422$& $\bar{1}$ \\ 
                 & & $4/m'mm$ & $4mm$& $\bar{1}$ \\ 
                 & & $4'/m'm'm$ & $\bar{4}2m$& $\bar{1}$ \\ 
                 $4'/mmm'$ & $(\times,\times)$ & $4'22'$ & $222$& $2_\perp$\\ 
                 & & $4'mm'$ & $2mm$& $2_\perp$ \\ 
                 & & $\bar{4}'2m'$ & $222$& $m_\perp$ \\ 
                 & & $\bar{4}'2'm$ & $2mm$& $m_\perp$ \\ 
                 & & $4'/mmm'$ & $mmm$& $2_\perp$ \\ 
                 $4/mm'm'$ & $(\times,\checkmark)$ & $4m'm'$ & $4$& $m_\perp$\\ 
                 & & $42'2'$ & $4$& $2_\perp$ \\ 
                 & & $\bar{4}2'm'$ & $\bar{4}$& $m_\perp$ \\ 
                 & & $4/mm'm'$ & $4/m$& $2_\perp$ \\ 
                 $6/mmm1'$ & $(\times,\times)$ & $6221'$ & $622$& $1$ \\ 
                 & & $6mm1'$ & $6mm$& $1$ \\ 
                 & & $\bar{6}2m1'$ & $\bar{6}2m$& $1$ \\ 
                 & & $6/mmm1'$ & $6/mmm$& $1$ \\ 
                 & & $6'/mmm'$ & $\bar{6}m2$& $\bar{1}$ \\ 
                 & & $6/m'm'm'$ & $622$& $\bar{1}$ \\ 
                 & & $6/m'mm$ & $6mm$& $\bar{1}$ \\ 
                 $6/mm'm'$ & $(\times ,\checkmark)$ & $\bar{6}m'2'$ & $\bar{6}$& $2_\perp$ \\ 
                 & & $\bar{6}'m'2$ & $32$& $m_\perp$ \\ 
                 & & $62'2'$ & $6$& $2_\perp$ \\ 
                 & & $6m'm'$ & $6$& $m_\perp$ \\ 
                 & & $6/mm'm'$ & $6/m$& $m_\perp$ \\ 
                 $6'/m'm'm$ & $(\times,\times)$ & $\bar{6}'m2'$ & $3m$& $2_\perp$ \\ 
                 & & $6'2'2$ & $32$& $2_\perp$ \\ 
                 & & $6'm'm$ & $32$& $m_\perp$ \\ 
                 & & $6'/m'm'm$ & $\bar{3}m$& $2$ \\ 
                 \bottomrule
\end{longtable}

\section{Odd-parity cross-circular Raman optical activity}
\label{Sec_odd_parity_CCROA}

The odd-parity CCROA (CC$_-$) undergoes the \Pa{}-operation transformation as
                \begin{equation}
                        \text{CC}_- (\omega,\delta \omega ,\system) = - \text{CC}_- (\omega,\delta \omega ,I \system I^{-1}).
                        \label{CCROA_odd_inversion_constraint}
                \end{equation}
Combining the constraint from the out-of-plane mirror reflection $m_\perp$, the odd-parity CCROA does not differ in systems with $\system$ and with its rotation image $2_\perp \system 2_\perp^{-1}$ ($2_\perp = I \cdot m_\perp $);
                \begin{equation}
                        \text{CC}_- (\omega,\delta \omega ,\system) = \text{CC}_- (\omega,\delta \omega ,2_\perp \system 2_\perp^{-1}).
                \end{equation}
In parallel with discussions for the even-parity CCROA, let us make use of the class of the point groups merged by the out-of-plane two-fold rotation.

For instance, the magnetic point groups for trigonal systems are classified into the five classes
        \begin{align}
        &32 = \left\{ 3,32\right\},\\
        &\bar{3}m = \left\{ 3m,\bar{3},\bar{3}m\right\},\\
        &321' = \left\{ 31',321',32'\right\},\\
        &\bar{3}m1' = \left\{ 3m1',\bar{3}1',\bar{3}m1', \bar{3}m',\bar{3}'m \right\},\\
        &\bar{3}'m' = \left\{ \bar{3}', 3m', \bar{3}'m'\right\}.
        \end{align}
The obtained classes are as follows.
The chiral class $321'$ has the \T{} symmetry but no improper rotation and is characterized by the \T{}-even pseudo-scalar~\cite{Barron1986-yy}.
The black-white class $\bar{3}'m'$ also has no improper rotation in its unitary operations but conserves the \PT{} symmetry which forbids the \T{}-even pseudo-scalar.
On the other hand, the \T{}-odd and \PT{}-even pseudo-scalar, namely magnetic chirality, is allowed in $\bar{3}'m'$.
The chiral class $321'$ is characterized by CC$_-^\text{\, s}$, while the $\bar{3}'m'$ is the magnetic chiral class allows for only CC$_-^\text{\, a}$ because the preserved \PT{} symmetry forbids the \T{}-even CCROA (CC$_-^\text{\, s}$).
Two types of odd-parity CCROA are concurrently allowed in the colorless class $32$, which shows the chirality as well as the magnetic chirality and hence can be called the composite-chiral class.

CCROA for the chiral class has been confirmed in systems with chiral motifs such as the chiral solids and molecules~\cite{Nafie1989-hj,Che1991-vn,Vargek1997-rv,Ishito2022-fr,Oishi2024-dy}.
The Stokes and anti-Stokes parts of CCROA observed in \cite{Ishito2022-fr,Oishi2024-dy} are consistent with the formula for CC$_-^\text{\, s}$~\cite{Barron1976-hs,Barron1985-mg}.
On the other hand, they show the opposite signs in materials belonging to the magnetic chiral class such as the \PT{}-symmetric magnets, Cr$_2$O$_3$, even-layered MnBi$_2$Te$_4$, and so on~\cite{Watanabe2024-hw}.

In contrast to the noncentrosymmetric classes, the class $\bar{3}m1'$ is achiral and shows \T{} symmetry and improper rotation symmetries.
Finally, the colorless class $\bar{3}m$ similarly shows the improper rotation symmetry without any antiunitary symmetry such as \T{}, implying the magnetic achiral symmetry. 
The achiral and magneto-achiral classes do not show the odd-parity CCROA due to their $m_\perp$ or $I$ symmetry.

We show the classification table of CC$_-$ in Table~\ref{Table_odd_parity_CCROA_BW_chiral-class} for the non-cubic magnetic point groups.
Supposing that the symmetry of a magnetic point group is enhanced as long as its holohedry is maintained~\cite{Aroyo2016-mg}, we obtained each class by using the $n$-fold ($g= n$) and two-fold ($g = 2_\perp$) rotations which are parallel and perpendicular to the principal axis, respectively.

We also show the summary of CCROA and PCROA in Fig.~\ref{Fig_motifs}.
The response is divided into four types in terms of the parity under the \Pa{} and \T{} operations.
Note that, similarly to the case of CCROA, PCROA is classified based on the magnetic Laue and a series of chiral and achiral classes.
Different types of ROA can be admixed with each other in those with significant symmetry violations; \textit{e.g.,} in the case of the \T{}-even CCROA, a chiral and ferroaxial crystal can host both of CC$_+^\text{\, s}$ and CC$_-^\text{\, s}$~\cite{Martinez2024-nd}.

\begin{longtable}{cccccc}
        \caption{
                Classification of odd-parity CCROA (CC$_-$) by achiral (aC), chiral (C), magneto-achiral (m-aC), magnetic chiral (m-C), and composite-chiral (c-C) classes.
                Each class is labeled by the \T{}-even and \T{}-odd odd-parity CCROA; \textit{e.g.}, $(\times, \checkmark)$ denotes no \T{}-even CCROA, but \T{}-odd CCROA allowed.
                The notations for $M$, $G$, $g$ are same as those in Table~\ref{Table_magnetic-Laue-class}.
                }
        \label{Table_odd_parity_CCROA_BW_chiral-class}\\
        
        \toprule
        Class & Type & CCROA & $M$ & $G$ & $g$ \\
        \midrule
        \endfirsthead
        
        \multicolumn{6}{c}{\tablename\ \thetable\ (\textit{cont.})} \\
        Class & Type& CCROA & $M$ & $G$ & $g$ \\
        \midrule
        \endhead
        
        \endfoot
        
        \endlastfoot

        $\bar{1}$ & m-aC& $(\times,\times)$ & $\bar{1}$ & $\bar{1}$& \\ 
        $1$ & c-C& $(\checkmark,\checkmark)^\ast$ & $1$ & $1$& \\ 
        $2/m$ & m-aC & $(\times,\times)$  & $m$ & $m$& \\ 
        &&& $2/m$ & $2/m$& \\ 
        $2$ & c-C & $(\checkmark,\checkmark)^\ast$& $2$ & $2$& \\ 
        $mmm$ & m-aC & $(\times,\times)$& $mmm$ & $mmm$& \\ 
        &&& $2mm$ & $2mm$& \\ 
        $222$ & c-C & $(\checkmark,\checkmark)^\ast$& $222$ & $222$& \\ 
        $\bar{3}m$ & m-aC & $(\times,\times)$& $\bar{3}$ & $\bar{3}$& \\ 
        &&& $3m$ & $3m$& \\ 
        &&& $\bar{3}m$ & $\bar{3}m$& \\ 
        $32$ & c-C & $(\checkmark,\checkmark)$& $3$ & $3$& \\ 
        &&& $32$ & $32$& \\ 
        $4/mmm$ & m-aC & $(\times,\times)$& $\bar{4}$ & $\bar{4}$& \\ 
        &&& $4/m$ & $4/m$& \\ 
        &&& $4mm$ & $4mm$& \\ 
        &&& $\bar{4}2m$ & $\bar{4}2m$& \\ 
        &&& $4/mmm$ & $4/mmm$& \\ 
        $422$ & c-C & $(\checkmark,\checkmark)$& $4$ & $4$& \\ 
        &&& $422$ & $422$& \\ 
        $6/mmm$ & m-aC & $(\times,\times)$& $\bar{6}$ & $\bar{6}$& \\ 
        &&& $6/m$ & $6/m$& \\ 
        &&& $6mm$ & $6mm$& \\ 
        &&& $\bar{6}2m$ & $\bar{6}2m$& \\ 
        &&& $6/mmm$ & $6/mmm$& \\ 
        $622$ & c-C & $(\checkmark,\checkmark)$& $6$ & $6$& \\ 
        &&& $622$ & $622$& \\ 
        \midrule
        $1'$ &C& $(\checkmark,\times)$&$1'$ & $1$&$1$ \\ 
        $\bar{1}1'$ &aC& $(\times,\times)$&$\bar{1}1'$ & $\bar{1}$&$1$ \\ 
        $\bar{1}'$ &m-C& $(\times,\checkmark)$& $\bar{1}'$ & $1$&$\bar{1}$ \\ 
        $2/m1'$ &aC& $(\times,\times)$& $2/m1'$ & $2/m$& $1$\\ 
        & & & $m1'$ & $m$&$1$ \\ 
        & & & $2/m'$ & $\bar{1}$&$2$ \\ 
        $21'$ &C& $(\checkmark,\times)$& $21'$ & $2$& $1$\\ 
        & & & $2'$ & $1$&$2$ \\ 
        $2/m'$ &m-C& $(\times,\checkmark)$&$2/m'$ & $2$& $\bar{1}$\\ 
        & & & $m'$ & $1$&$m$ \\ 
        $mmm1'$ &aC& $(\times,\times)$& $mmm1'$ & $mmm$&$1$ \\ 
        & & & $2mm1'$ & $2mm$&$1$ \\ 
        & & & $m'm2'$ & $m$&$2$ \\ 
        & & & $m'mm$ & $2mm$&$\bar{1}$ \\ 
        & & & $m'm'm$ & $2/m$&$2$ \\ 
        $2221'$ &C& $(\checkmark,\times)$& $2221'$ & $222$&$1$ \\ 
        & & & $2'2'2$ & $2$&$2$ \\ 
        $m'm'm'$ &m-C& $(\times,\checkmark)$& $m'm'm'$ & $222$&$\bar{1}$ \\ 
        & & & $m'm'2$ & $2$&$m$ \\ 
        $\bar{3}m1'$ &aC& $(\times,\times)$& $\bar{3}m1'$ & $\bar{3}m$& $1$\\ 
        & & & $\bar{3}1'$ & $\bar{3}$&$1$ \\ 
        & & & $3m1'$ & $3m$&$1$ \\ 
        & & & $\bar{3}'m$ & $3m$&$\bar{1}$ \\ 
        & & & $\bar{3}m'$ & $\bar{3}$&$m$ \\ 
        $321'$ &C& $(\checkmark,\times)$& $321'$ & $32$& $1$\\ 
        & & & $31'$ & $3$&$1$ \\ 
        & & & $32'$ & $3$&$2$ \\ 
        $\bar{3}'m'$ &m-C& $(\times,\checkmark)$& $\bar{3}'m''$ & $32$& $\bar{1}$\\ 
        & & & $\bar{3}'$ & $3$&$\bar{1}$ \\ 
        & & & $3m'$ & $3$&$m$ \\ 
        $4/mmm1'$ &aC& $(\times,\times)$& $4'/m$ & $2/m$& $\bar{4}$\\ 
        &&& $4'/m'$ & $\bar{4}$& $\bar{1}$\\ 
        &&& $4/m1'$ & $4/m$& $1$\\ 
        &&& $\bar{4}1'$ & $\bar{4}$& $1$\\ 
        &&& $\bar{4}2'm'$ & $\bar{4}$& $m_\perp$ \\ 
        &&& $4/mm'm'$ & $4/m$& $2_\perp$ \\ 
        &&& $4'/mmm'$ & $mmm$& $4$ \\ 
        &&& $\bar{4}'2'm$ & $2mm$& $\bar{4}$ \\ 
        &&& $4'mm'$ & $2mm$& $4$ \\ 
        &&& $\bar{4}2m1'$ & $\bar{4}2m$&$1$ \\ 
        &&& $4mm1'$ & $4mm$& $1$\\ 
        &&& $4/mmm1'$ & $4/mmm$& $1$ \\ 
        &&& $4/m'mm$ & $4mm$& $\bar{1}$ \\ 
        &&& $4'/m'm'm$ & $\bar{4}2m$& $\bar{1}$ \\ 
        $4221'$ &C& $(\checkmark,\times)$& $4'$ & $2$& $4$\\ 
        &&& $41'$ & $4$& $1$\\ 
        &&& $4'22'$ & $222$& $4$\\ 
        &&& $42'2'$ & $4$& $2_\perp$ \\ 
        &&& $4221'$ & $422$& $1$\\ 
        $4/m'm'm'$ &m-C&$(\times,\checkmark)$& $\bar{4}'$ & $2$& $\bar{4}$\\ 
        &&& $4/m'$ & $4$& $\bar{1}$\\ 
        &&& $4m'm'$ & $4$& $m_\perp$\\ 
        &&& $\bar{4}'2m'$ & $222$& $\bar{4}$ \\ 
        &&& $4/m'm'm'$ & $422$& $\bar{1}$ \\ 
        $6/mmm1'$ &aC& $(\times,\times)$& $\bar{6}1'$ & $\bar{6}$& $1$\\ 
        &&& $6/m1'$ & $6/m$& $1$\\ 
        &&& $6'/m$ & $\bar{6}$& $\bar{1}$\\ 
        &&& $6'/m'$ & $\bar{3}$& $6$\\ 
        &&& $6mm1'$ & $6mm$& $1$ \\ 
        &&& $\bar{6}2m1'$ & $\bar{6}2m$& $1$ \\ 
        &&& $6/mmm1'$ & $6/mmm$& $1$ \\ 
        &&& $6'/mmm'$ & $\bar{6}m2$& $\bar{1}$ \\ 
        &&& $6/m'mm$ & $6mm$& $\bar{1}$ \\ 
        &&& $\bar{6}m'2'$ & $\bar{6}$& $2_\perp$ \\ 
        &&& $6/mm'm'$ & $6/m$& $m_\perp$ \\ 
        &&& $\bar{6}'m2'$ & $3m$& $2_\perp$ \\ 
        &&& $6'/m'm'm$ & $\bar{3}m$& $2$ \\ 
        $6221'$ &C& $(\checkmark,\times)$& $61'$ & $6$& $1$\\ 
        &&& $6'$ & $3$& $2$\\ 
        &&& $6221'$ & $622$& $1$ \\ 
        &&& $62'2'$ & $6$& $2_\perp$ \\ 
        &&& $6'2'2$ & $32$& $2_\perp$ \\ 
        $6/m'm'm'$ &m-C& $(\times,\checkmark)$& $6/m'$ & $6$& $\bar{1}$\\ 
        &&& $\bar{6}'$ & $3$& $\bar{6}$\\ 
        &&& $6/m'm'm'$ & $622$& $\bar{1}$ \\ 
        &&& $\bar{6}'m'2$ & $32$& $m_\perp$ \\ 
        &&& $6m'm'$ & $6$& $m_\perp$ \\ 
        &&& $6'm'm$ & $32$& $m_\perp$ \\ 
        \bottomrule
\end{longtable}

\expandafter\ifx\csname iffigure\endcsname\relax
                \begin{figure*}[htbp]
                \centering
                \includegraphics[width=0.9\linewidth,clip]{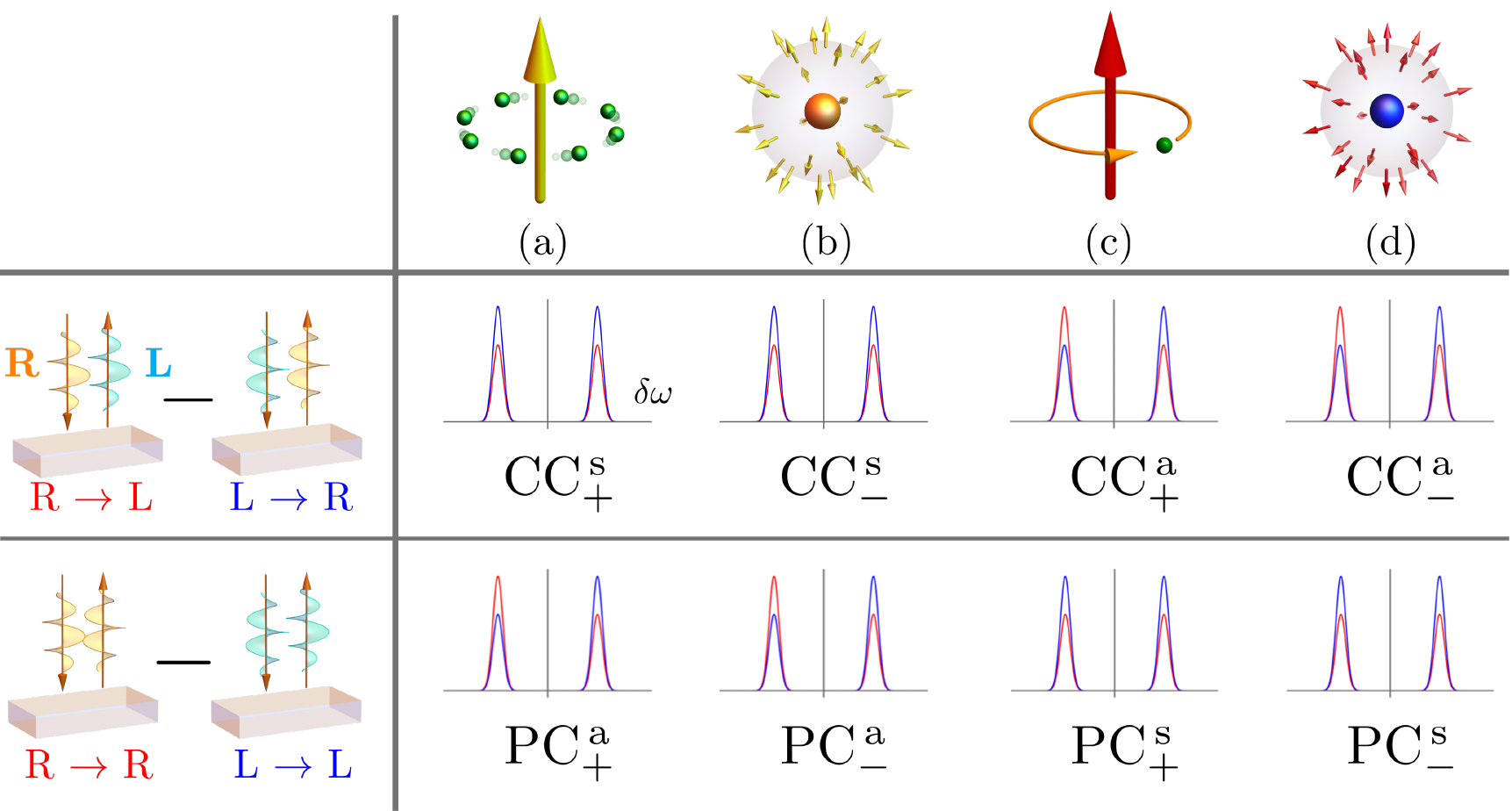}
                \caption{
                        Motifs of materials with the cross-circular Raman optical activity (CCROA) denoted by CC$_X^{\,Y}$ and parallel-circular Raman optical activity (PCROA) denoted by PC$_X^{\,Y}$ ($X=\pm$, $Y=\text{s},\text{a}$).
                        ROA is classified by the space-inversion ($I$) and time-reversal ($\theta$) parities.
                        For instance, the \T{}-even and \T{}-odd parts respectively point to the variations of the left-handed circularly-polarized light response to the right-handed circularly-polarized light (R$\to$L, colored in red) and the right-handed one to the left-handed stimulus (L$\to$R, colored in blue) in the identical and staggered manner between Stokes ($\delta \omega <0$) and anti-Stokes ($\delta \omega >0$) peaks.
                        Note that we suppress the difference stemming from the Bose-Einstein distribution function (Eq.~\eqref{CCROA_time-reversal}).
                        (a) The even-parity and \T{}-even part (CC$_+^\text{\, s}$, PC$_+^\text{\, a}$) corresponds to the ferroaxial motif consisting of the circulating electric polarizations (green-colored spheres are displacing electrons, yellow arrow denotes the ferroaxial vector $\bm{A}$). 
                        (b) The odd-parity and \T{}-even part (CC$_-^\text{\, s}$, PC$_-^\text{\, a}$) for the chiral motif formed by source or sink of ferroaxial vectors as $\nabla \cdot \bm{A}$~\cite{Oiwa2022-vu,Oiwa2025-ar}. 
                        (c) The even-parity and \T{}-odd part (CC$_+^\text{\, a}$, PC$_+^\text{\, s}$) for the magneto-axial motif such as magnetization $\bm{M}$ (red-colored arrow) associated with the circulating electric current colored in orange. 
                        (d) The odd-parity and \T{}-odd part (CC$_-^\text{\, a}$, PC$_-^\text{\, s}$) for the magnetic chiral motif represented by source or sink of magnetization, $\nabla \cdot \bm{M}$.
                        }
                \label{Fig_motifs}
                \end{figure*}
\fi

\section{CCROA and PCROA of cubic systems}
\label{Sec_cubic_ROA}

We do not make a symmetry analysis for the cubic systems in the previous sections because the primary direction cannot be chosen without ambiguity.
Let us take several incident directions and consider the symmetry of CCROA in the normal reflection geometry.
The symmetry of PCROA can be derived from that of CCROA.

To prevent the response from being admixed with the birefringence effect, the incident direction $\bm{q}_\text{i}$ is taken to be the high symmetry rotation axes with the $n$-fold rotation symmetry ($n\geq 3$).
The candidates are the $[001]$, $[111]$, and their crystallographically equivalent axes of the cubic systems in the conventional setting.
We consider the even-parity and odd-parity CCROA for the $[001]$ and $[111]$ incident directions as follows.

In the case of $\bm{q}_\text{i} = [001]$ incidence, CCROA is forbidden in the presence of the $m_\perp$ symmetry whose mirror plane contains the $[001]$ axis.
Similarly to the case of non-cubic systems, the even-parity CCROA is classified by the magnetic Laue class for cubic point groups.
We tabulate these classes as
                \begin{equation}
                m\bar{3}, m\bar{3}m,m\bar{3}1', m\bar{3}m1', m\bar{3}m'.
                \label{cubic_magnetic_laue_class}
                \end{equation}
Every class contains the $m_\perp$ or $2_\perp = I m_\perp$ symmetry forbidding CCROA regarding the $[001]$ incidence.
It follows that all the classes in Eq.~\eqref{cubic_magnetic_laue_class} do not allow for the even-parity CCROA.

The odd-parity CCROA for $\bm{q}_\text{i} = [001]$ is classified by the class enhanced by the two-fold rotation perpendicular to $[001]$ or by $n$-fold rotation parallel to $[001]$.
The obtained classes are
                \begin{equation}
                m\bar{3}m,432, m\bar{3}m1',4321', m'\bar{3}'m',
                \label{cubic_holohedry_class}        
                \end{equation}
among which the composite-chiral ($432$), chiral ($4321'$), and magnetic chiral ($m'\bar{3}'m'$) cases imply CCROA.

Next, the incident direction is taken to be $\bm{q}_\text{i} = [111]$.
The odd-parity CCROA is similarly classified as that with $\bm{q}_\text{i} = [001]$.
After the enhancement of point groups with respect to the $[111]$ direction, we obtain the same series of classes tabulated in Eq.~\eqref{cubic_holohedry_class}.
On the other hand, the even-parity CCROA for $\bm{q}_\text{i} = [111]$ differs from that for $\bm{q}_\text{i} = [001]$ where any magnetic Laue class of Eq.~\eqref{cubic_magnetic_laue_class} does not possess the even-parity CCROA.
The $m_\perp$ symmetry for the $[111]$ incidence is not present in the colorless ($m\bar{3}$), gray ($m\bar{3}1'$), and black-white ($m\bar{3}m'$) classes, and thereby CCROA is allowed in spite of neither ferroaxial nor magneto-axial anisotropy.

The obtained classification is summarized in Table~\ref{Table_magnetic-Laue-class_cubic} for the even-parity case and Table~\ref{Table_odd_parity_CCROA_BW_chiral-class_cubic} for the odd-parity case.
Although we presented the symmetry analysis of CCROA, the classification for PCROA is the same as that for CCROA.
The detailed analysis with the microscopic calculations is presented elsewhere~\cite{watanabe_multiaxial}.

\begin{longtable}{cccccc}
        \caption{
                Classification of even-parity cross- and parallel-circular Raman optical activities (CC$_+$, PC$_+$) for the cubic systems.
                Each class is labeled by the \T{}-even and \T{}-odd ROA regarding the $[001]$ and $[111]$ incidence; \textit{e.g.}, ROA$_{[001]}=(\times, \checkmark)$ denotes no \T{}-even ROA but allowed \T{}-odd ROA for the incident light with $\bm{q}_\text{i} =[001]$.
                Other notations are the same as those given in Table~\ref{Table_magnetic-Laue-class}.
                }
        \label{Table_magnetic-Laue-class_cubic}\\
        
        \toprule
        Laue class & ROA$_{[001]}$ &ROA$_{[111]}$ & $M$ & $G$ & $g$ \\
        \midrule
        \endfirsthead
        
        \multicolumn{6}{c}{\tablename\ \thetable\ (\textit{cont.})} \\
        Laue class & ROA$_{[001]}$ &ROA$_{[111]}$ & $M$ & $G$ & $g$ \\
        \midrule
        \endhead
        
        \endfoot
        
        \endlastfoot
                 $m\bar{3}m$ & $(\times,\times)$ &$(\times,\times)$ & $\bar{4}3m$ & $\bar{4}3m$& \\ 
                 &&& $432$ & $432$& \\ 
                 &&& $m\bar{3}m$ & $m\bar{3}m$& \\
                 $m\bar{3}$ & $(\times,\times)$ & $(\checkmark,\checkmark)$ & $23$ & $23$& \\ 
                 &&& $m\bar{3}$ & $m\bar{3}$& \\ 
                \midrule
                $m\bar{3}m1'$ & $(\times,\times)$ & $(\times,\times)$ & $\bar{4}3m1'$ & $\bar{4}3m$& $1$ \\ 
                &&& $4321'$ & $432$& $1$ \\ 
                &&& $m\bar{3}m1'$ & $m\bar{3}m$& $1$ \\
                &&& $m'\bar{3}m'$ & $432$& $\bar{1}$ \\
                &&& $m'\bar{3}m$ & $\bar{4}3m$& $\bar{1}$ \\
                $m\bar{3}1'$ & $(\times,\times)$& $(\checkmark,\times)$ & $231'$ & $23$& $1$ \\ 
                 &&& $m\bar{3}1'$ & $m\bar{3}$& $1$ \\ 
                 &&& $m'\bar{3}'$ & $23$& $\bar{1}$ \\ 
                 $m\bar{3}m'$ & $(\times,\times)$ & $(\times,\checkmark)$ & $\bar{4}'3m'$ & $23$& $m$ \\ 
                 &&& $4'32'$ & $23$& $4$ \\
                 &&& $m\bar{3}m'$ & $m\bar{3}$& $4$ \\
                 \bottomrule
\end{longtable}

\begin{longtable}{cccccc}
        \caption{
                Classification of odd-parity cross- and parallel-circular Raman optical activities (CC$_-$, PC$_-$) for the cubic systems.
                Each class is labeled by the \T{}-even and \T{}-odd ROA regarding the $[001]$ and $[111]$ incidence; \textit{e.g.}, ROA$=(\times, \checkmark)$ denotes no \T{}-even CCROA but allowed \T{}-odd ROA.
                Other notations are the same as those in Table~\ref{Table_odd_parity_CCROA_BW_chiral-class}.
                }
        \label{Table_odd_parity_CCROA_BW_chiral-class_cubic}\\
        
        \toprule
        Laue class & Type &ROA & $M$ & $G$ & $g$ \\
        \midrule
        \endfirsthead
        
        \multicolumn{6}{c}{\tablename\ \thetable\ (\textit{cont.})} \\
        Laue class & Type &ROA & $M$ & $G$ & $g$ \\
        \midrule
        \endhead
        
        \endfoot
        
        \endlastfoot
        $m\bar{3}m$ &m-aC& $(\times,\times)$&$m\bar{3}$ & $m\bar{3}$&  \\ 
        &&& $\bar{4}3m$ & $\bar{4}3m$&  \\ 
        &&& $m\bar{3}m$ & $m\bar{3}m$&  \\
        $432$ &c-C& $(\checkmark,\times)$& $23$ & $23$&  \\ 
        &&& $432$ & $432$&  \\ 
        \midrule
        $m\bar{3}m1'$ &aC& $(\times,\times)$&$m\bar{3}1'$ & $m\bar{3}$& $1$ \\ 
        &&& $\bar{4}3m1'$ & $\bar{4}3m$& $1$ \\ 
        &&& $m\bar{3}m1'$ & $m\bar{3}m$& $1$ \\
        &&& $m'\bar{3}m$ & $\bar{4}3m$& $\bar{1}$ \\
        &&& $m\bar{3}m'$ & $m\bar{3}$& $4$ \\
        $4321'$ &C& $(\checkmark,\times)$& $231'$ & $23$& $1$ \\ 
        &&& $4321'$ & $432$& $1$ \\ 
        &&& $4'32'$ & $23$& $4$ \\
        $m'\bar{3}'m'$ &mC& $(\times,\checkmark)$& $m'\bar{3}'$ & $23$& $\bar{1}$ \\ 
        &&& $m'\bar{3}m'$ & $432$& $\bar{1}$ \\
        &&& $\bar{4}'3m'$ & $23$& $m$ \\ 
        \bottomrule
\end{longtable}

\section{Discussion}
\label{Sec_summary}

In this work, we investigated the symmetry of the cross- and parallel-circular Raman optical activities and presented the systematic classification.
The classification is obtained by the magnetic Laue class for the even-parity case and the (magneto) achiral/chiral crystal class for the odd-parity case. 
The result not only covers the cases identified in prior works on the ferroaxial, magneto-axial, and chiral materials but also points out more diverse ROA, such as that from the magnetic chiral anisotropy.
Moreover, the formulation is based on the phenomenological arguments applicable to various kinds of Raman-scattering processes, such as spontaneous and coherent stimulated Raman scatterings.

It is noteworthy that the antiunitary symmetry of solids is closely tied to the relation between the Stokes and anti-Stokes signals.
Thus, the circular dichroism in the Raman scattering provides us with a powerful diagnosis of the symmetry of the target materials~\cite{Kung2015-fw,Kung2016-sw}.
For example, it may be feasible to identify the antiunitary symmetry of the exotic quantum phase implied by the measurement of bulk properties~\cite{Zhao2017-ri,MurayamaIshida2021}.
The useful property is in stark contrast to that of nonlinear optical responses such as photocurrent generation, in which the different antiunitary symmetry (\textit{e.g.}, \T{} or \PT{} symmetry) does not give qualitative differences while it implies the distinct mechanism for the response. 

Tables~\ref{Table_magnetic-Laue-class} and \ref{Table_odd_parity_CCROA_BW_chiral-class} also show that both of the \T{}-even and \T{}-odd ROA can occur in a certain case such as those belonging to the colorless and axial Laue class (\textit{e.g.,} $M = \bar{6}$).
It implies an intriguing situation that ROA is found in only the Stokes or anti-Stokes peak since the \T{}-even and \T{}-odd parts contribute to those peaks in a constructive or destructive manner.
Such `perfect' CCROA and PCROA may be realized in the ferroaxial materials under large magnetic fields or by combined with the magneto-axial order (\textit{e.g.}, ferromagnetic order).
The phenomenon may be feasible for the odd-parity case (see the composite-chiral class in Table~\ref{Table_odd_parity_CCROA_BW_chiral-class}).

The quantitative aspects of responses are an important issue to be addressed.
We expect that our symmetry analysis is convenient for exploring significant CCROA and PCROA with those microscopic calculations.

\section*{acknowledgement}
H.W. is grateful to Eiichi Oishi for helpful discussions.
R.O. was supported by Special Postdoctoral Researcher Program at RIKEN.
This work is supported by Grant-in-Aid for Scientific Research from JSPS KAKENHI Grant
No.~JP23K13058 (H.W.),
No.~JP24K00581 (H.W.),
No.~JP25H02115 (H.W.),
No.~JP21H04990 (R.A.),
No.~JP25H01246 (R.A.),
No.~JP25H01252 (R.A.),
JST-CREST No.~JPMJCR23O4(R.A.),
JST-ASPIRE No.~JPMJAP2317 (R.A.),
JST-Mirai No.~JPMJMI20A1 (R.A.),
MEXT X-NICS No.~JPJ011438 (T.S.),
NINS OML Project No.~OML012301 (T.S.),
JST CREST No.~JPMJCR24R5 (T.S.),
and RIKEN TRIP initiative (RIKEN Quantum, Advanced General Intelligence for Science Program, Many-body Electron Systems).
H.W. was also supported by JSR Corporation via JSR-UTokyo Collaboration Hub, CURIE.

\end{document}